# High Breakdown Field Multi-kV UWBG AlGaN Transistors

Seungheon Shin[1,a)], Kyle Liddy[3], Jon Pratt[1], Can Cao[1], Yinxuan Zhu[1], Brianna A. Klein[4], Andrew Armstrong[4], Andrew A. Allerman[4], and Siddharth Rajan[1,2]

[1]*Department of Electrical & Computer Engineering, The Ohio State University, Columbus OH 43210, USA*
[2]*Department of Materials Science & Engineering, The Ohio State University, Columbus OH 43210, USA*
[3]*Air Force Research Laboratory, Information and Spectrum Warfare Directorate, Wright-Patterson AFB, OH 45433, USA*
[4]*Sandia National Laboratories, Albuquerque, New Mexico 87123, USA*

**Abstract—** We demonstrate high-performance UWBG AlGaN PolFETs exhibiting a state-of-the-art combination of nearly 1 A/mm on-state current (~ 960 mA/mm) and large breakdown field (> 4.8 MV/cm) in high carrier density ($1.15 \times 10^{13}$ cm$^{-2}$). Multi-kV robustness is successfully demonstrated exhibiting 1.28 and 2.17 kV by utilizing a gate-connected field plate structures in 3.9 and 6.8 μm $L_{GD}$, corresponding to the extremely low specific on-resistance of 1.25 and 2.86 mΩ·cm$^2$, respectively. High RF performance is also achieved, providing $f_T$ and $f_{MAX}$, of 8.5 and 15 GHz, respectively, for 3.9 μm $L_{GD}$. These results highlight UWBG AlGaN as a platform for both high-voltage RF and power applications.

[a)] Authors to whom correspondence should be addressed
Electronic mail: *shin.928@osu.edu*

Ultra-wide-bandgap (UWBG) AlGaN transistors (channel Al composition > 40%) have attracted increasing interest for next-generation power switching and high-voltage RF electronics. This material offers access to large critical electric field (> 10 MV/cm), high saturation velocity (~ 2 × 10$^7$ cm/s), and high-frequency capability, enabling performance targets that are difficult to reach with conventional III-nitride transistor technologies. Recent demonstrations of UWBG AlGaN devices have reported breakdown fields above 5 MV/cm [23, 24, 47], high breakdown voltage [19, 21, 24, 47], and excellent high-frequency performance [1, 26] that highlight the potential of this material system for both Baliga's figure of merit (BFOM) in high-voltage switching and Johnson's figure of merit (JFOM) in RF power applications. Despite this strong material potential, converting these material properties into a practical transistor platform remains challenging. Prior UWBG AlGaN transistors have shown individual advances in breakdown field, kV-class blocking voltage, current drive, contact resistance, and RF performance. Even so, the simultaneous realization of multi-kV blocking, high saturated current density, and high MV/cm-class breakdown field has not yet been demonstrated. This performance gap has become one of the key issues for lateral UWBG AlGaN transistors intended for power conversion and high-voltage RF operation, where on-state conduction/saturation current and robust off-state field control must be achieved in the same device. Two device-level constraints are especially important in this regime. One is insufficient field management across the gate–drain region, which prevents the intrinsic critical field of UWBG AlGaN from being fully utilized in three-terminal operation, particularly as the gate–drain spacing is extended toward the multi-kV range. The other is the persistence of high access and contact resistance, which suppresses current drive and increases on-resistance. As shown in our work, contact formation in UWBG AlGaN HFETs remains nontrivial even when reverse-graded contact layers are used [23, 24], due to the additional tunneling obstacle induced by the barrier/channel heterostructure [24]. This tradeoff underscores the need for heterostructure design in UWBG AlGaN transistors that simultaneously optimizes lateral electrostatic field shaping and vertical carrier injection rather than improving either one in isolation.

In this work, we demonstrate UWBG AlGaN metal-insulator-semiconductor polarization-graded field-effect transistors (PolFETs) designed to address these coupled challenges. A PolFET removes the HFET barrier/channel vertical potential barrier for contacts. Compared to an HFET, the PolFET also offers lower sheet resistance ($R_{SH}$) by moving the channel into higher AlN mole-fraction regions with higher electron mobility ($\mu$). Higher channel mobility provides latitude in designing the charge density ($n_s$) while maintaining a target $R_{SH} = 1/q\mu n_s$. This added design freedom allows $n_s$ to be chosen to optimize lateral on-state conduction and lateral off-state field management simultaneously. The devices exhibit nearly 1 A/mm maximum on-state current together with a breakdown field of 4.8 MV/cm, while also achieving multi-kV breakdown voltage at longer gate–drain spacing with a low specific on-resistance. These results highlight UWBG AlGaN PolFETs as a promising candidate for lateral transistors requiring concurrent high current density, strong field control, and kV-class voltage handling for next-generation high power switching and high-voltage RF applications.

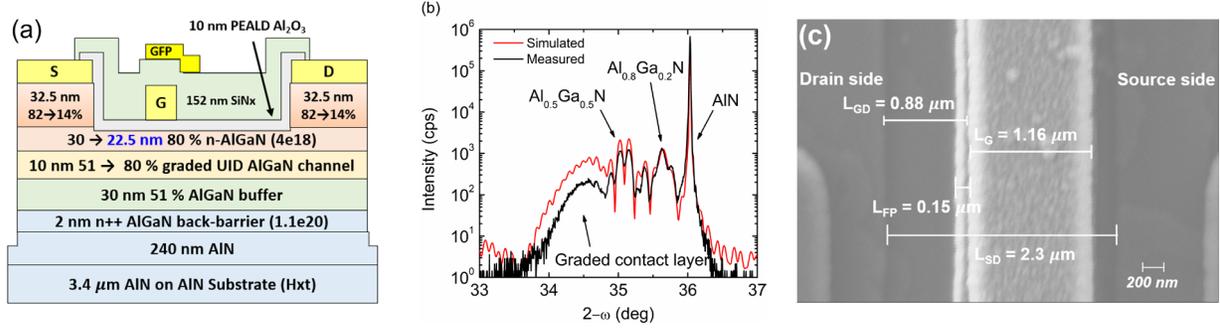

Figure 1. (a) Schematic of epitaxial structures and fabricated device structures for PolFETs, (b) high-resolution X-ray diffraction 2θ-ω scans (c) Top view SEM image for processed devices with $L_{GD}$ = 0.88 μm

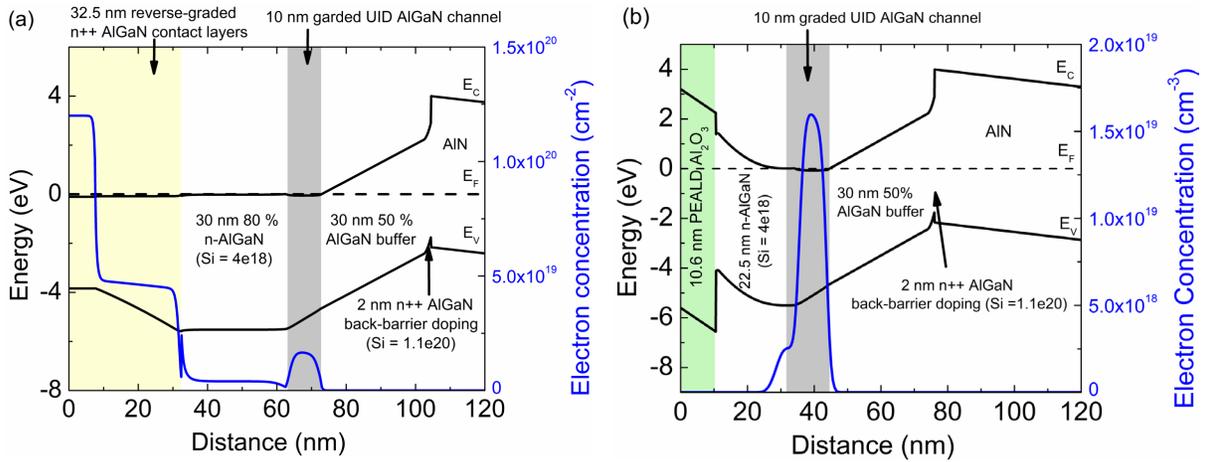

Figure 2. Simulated energy band diagrams and electron distributions (a) ohmic contact regions at equilibrium condition, cutline from the surface of reverse-graded n++ AlGaN contact layer to 50 % AlGaN buffer layer, (b) underneath the gate in PolFETs

A PolFET epitaxial layer structure was designed and grown on AlN substrates using a TNSC-4000HT metal-organic chemical vapor deposition (MOCVD) reactor (Fig. 1(a)). The epitaxial layers consist of a 2 nm of uniformly doped back-barrier layer (Si ~ 1.1 × $10^{20}$ cm$^{-3}$, net Si density ~ 2.2 × $10^{13}$ cm$^{-2}$) placed underneath 30 nm of 50 % unintentionally-doped (UID) AlGaN buffer layer to eliminate the parasitic hole gas in the back-barrier. A 10 nm UID graded channel layer (Al % from 50 to 80 % toward the top) was grown on top of the buffer layer, followed by a 30 nm n-AlGaN spacer layer (Si ~ 4 × $10^{18}$ cm$^{-3}$). Finally, 32.5 nm-thick reversed graded n++ AlGaN contact layers were formed on n-AlGaN spacer. The 30 nm n-AlGaN spacer serves two functions: (i) preventing 3DEG depletion from the surface, and (ii) acting as a sacrificial etching layer during removal of contact layers in the access and gate regions. The initial spacer layer thickness and doping concentration were designed based on energy band diagram simulations, considering both minimal surface depletion and extra thickness as an over-etching spacer. High-resolution X-ray diffraction (HR-XRD, Bruker D8 Discover) was employed to estimate Al composition for each layer as shown in Fig 1(b). Fig. 2(a) shows simulated energy band diagrams under the ohmic contact, degenerated Fermi level from contact layers to the UID PolFET channel, and calculated electron distribution profiles. Fig. 2(b) provides the 10.6 nm PEALD

Al$_2$O$_3$-integrated energy band diagram under the gate metal, confirming that the design minimize the surface and back depletion in the graded UID AlGaN channel layer and designed 3DEG density in this work.

For the device fabrication, direct-write optical lithography was used for all patterning steps. To define the source/drain spacings and access region, low-damage BCl$_3$/Cl$_2$/Ar (5/50/5 sccm) ICP-RIE etching was performed at ICP/RIE power = 40/8 W at 5 mTorr, targeting complete removal of reverse-graded contact layers by controlled over-etching into the n-AlGaN spacer. The barrier thickness after etching was inspected via atomic force microscopy and measured to be ~ 22.5 nm. Non-alloyed ohmic metal stacks (Ti/Al/Ni/Au = 20/120/30/100 nm) were formed using electron-beam (E-beam) evaporation. A 10.6 nm Al$_2$O$_3$ gate dielectric was then deposited using plasma-enhanced atomic layer deposition (Veeco Fiji G2 PE-ALD), with thickness confirmed on a bare Si control wafer using spectroscopic ellipsometry. 280 nm of mesa isolation was done using BOE wet etching for Al$_2$O$_3$ followed by ICP-RIE etch. Ni/Au/Ni (30/100/20 nm) gate metal was deposited via E-beam evaporation. Finally, the passivation with 152 nm PECVD SiN$_x$ and gate-connected field plate (GFP) (Ni/Au/Ni = 30/100/20 nm) was deposited. The schematic for the processed device is shown on Fig. 1(a). Fig. 1(c) presents scanning electron microscope image of a representative device, indicating $L_{SG}$ = 0.26 µm, $L_G$ = 1.16 µm, $L_{GD}$ = 0.88 µm and $L_{FP}$ = 0.15 µm. Other devices reported here have dimensions $L_{GD}$/$L_{FP}$ = 3.9/0.3 µm and 6.8/0.4 µm, with the same nominal $L_{SG}$ and $L_G$.

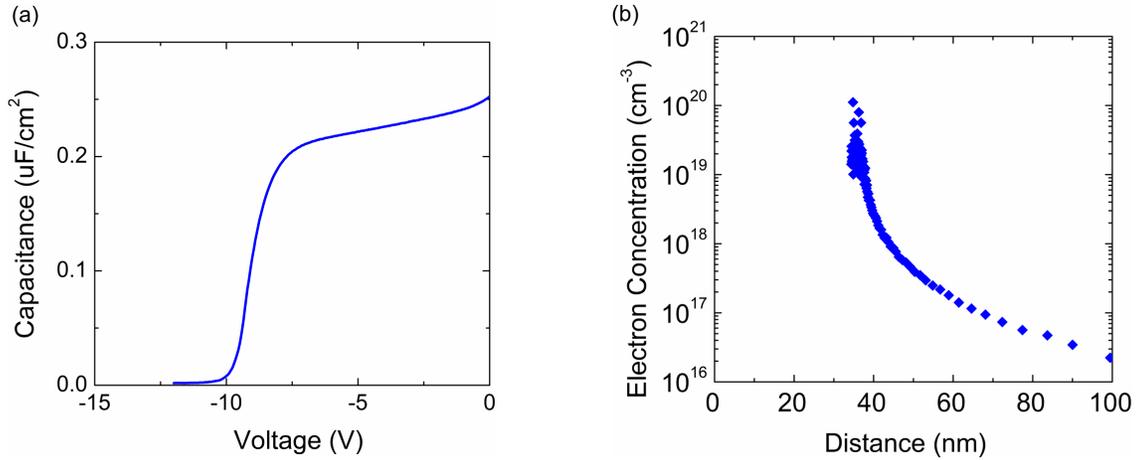

Figure 3. (a) C−V characteristics measured on MIS diode, (b) extracted electron distribution profile from C-V measurement

Transmission line measurements were done to investigate the ohmic contact properties, showing 2.07 Ω·mm contact resistance ($R_C$), 1.78 × 10$^{-5}$ Ω·cm$^2$ of contact resistivity ($\rho_C$), and 2.4 kΩ/□ of sheet resistance ($R_{SH}$). It should be noted that these $R_C$ and $\rho_C$ are relatively higher than those demonstrated in our previous work using similar epitaxial layers (~ 1 Ω·mm) [1]. This difference may be possibly due to unexpected variations in either ohmic process or epitaxial layers. Hall measurements indicate total sheet charge density of 1.15 × 10$^{13}$ cm$^{-2}$, $R_{SH}$ of 2.45 kΩ/□, and high electron mobility of 220 cm$^2$/V·s which results from the lack of ionized dopants in UID PolFET channel layer. To our best knowledge, this electron mobility is the highest reported electron mobility in UWBG AlGaN transistors.

The capacitance-voltage (C−V) characteristics were investigated to determine detailed 3DEG profile. The C−V measurements were conducted on MIS diode structures containing the 10.6 nm PEALD $Al_2O_3$ (Fig.3(a)). The extracted electron profile is presented in Fig. 3(b), confirming the role of the n-AlGaN spacer in preventing surface depletion and enabling high density of 3DEG formation. The estimated total charge density was ~ $1.26 \times 10^{13}$ cm$^{-2}$.

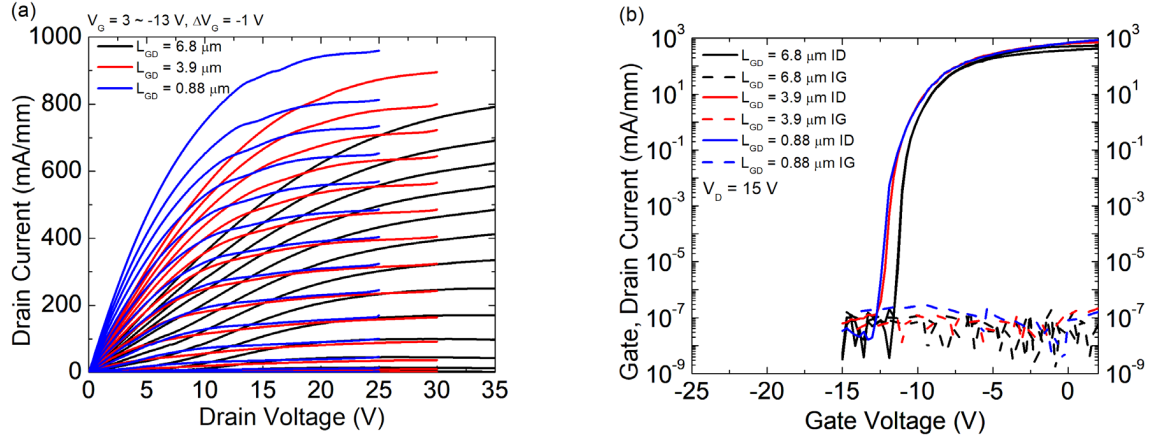

Figure 4. (a) Output curves at $V_{GS}$ = 3 ~ 13 V, $\Delta V_{GS}$ = -1, (b) transfer curves in log scale measured at $V_{DS}$ = 15 V, solid lines: $I_D$, dashed lines: $I_G$

DC current-voltage (I−V) characteristics were done using Keithley 4200. Overall devices showed pinch-off voltage ($V_P$) of − 12.5 V. A low gate leakage current (~ $2.2 \times 10^{-7}$ mA/mm) was consistently observed for devices with different $L_{GD}$, indicating a high $I_{ON}/I_{OFF}$ (> $4.8 \times 10^9$) due to the integration of PEALD $Al_2O_3$ gate dielectric (Fig 4(b)). Excellent maximum on-state current ($I_{MAX}$) of nearly 1 A/mm was measured, reaching 960, 900, and 800 mA/mm for $L_{GD}$ = 0.88, 3.9, and 6.8 μm, respectively (Fig. 4(a)). The estimated specific on-resistance ($R_{on.sp} = R_{ON} \times (2L_T + L_{SD})$) extracted from the linear region of DC output curves was 0.4, 1.25, and 2.86 mΩ·cm$^2$ for $L_{GD}$ = 0.88, 3.9, and 6.8 μm, respectively. Furthermore, the peak transconductance was between 78 ~ 100 mS/mm at $V_G$ ~ -2.25 V.

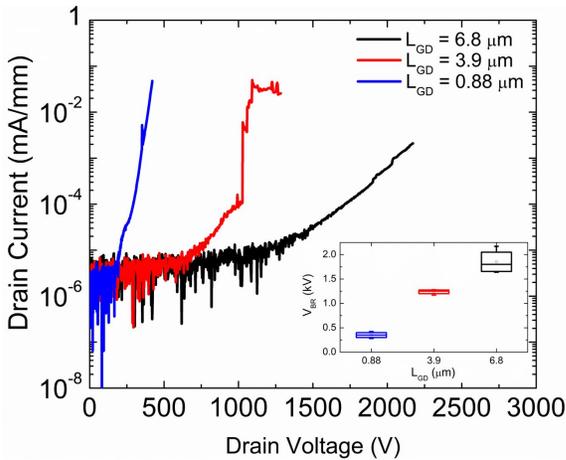

Figure 5. Three-terminal breakdown measurements for different $L_{GD}$ devices

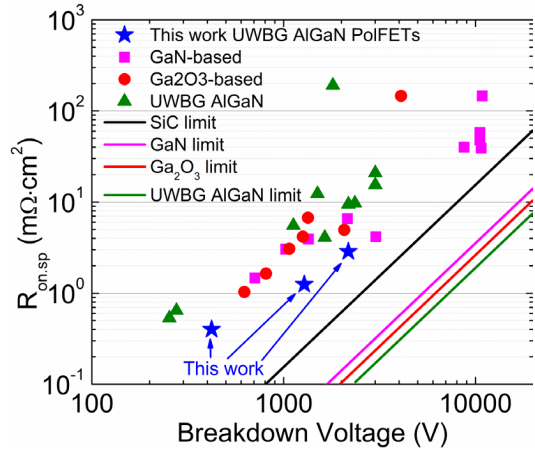

Figure 6. $R_{on.sp}$ vs. $V_{BR}$ Benchmark for various material platforms [2-24]

For breakdown evaluation, three-terminal high-voltage I−V measurements were conducted employing Keysight B1505A analyzer. The breakdown criterion was defined as the drain voltage point corresponding to 1 mA/mm leakage current level. The applied gate bias was $V_{GS} = V_p$ -5 V for each device. For $L_{GD} = 0.88$ μm device, breakdown voltage ($V_{BR}$) of 421 V was obtained, corresponding to high average breakdown field ($F_{BR} = V_{BR}/L_{GD}$) > 4.8 MV/cm (Fig. 5). Multi-kV breakdown robustness was achieved in devices with $L_{GD}$ = 3.9 and 6.8 μm, yielding 1.28 kV ($F_{BR}$ > 3.28 MV/cm) and 2.17 kV ($F_{BR}$ > 3.19 MV/cm), respectively. For $L_{GD} = 6.8$ μm devices, a low leakage current (~ 2 μA/mm) was maintained up to breakdown. The comparison of breakdown robustness and $L_{GD}$-dependent breakdown field trends between no field plate structures and GFP structures is presented in Fig. S1(a), (b). Based on breakdown measurements and $R_{on.sp}$, the estimated BFOM is ~ 1.65 GW/cm². These results place UWBG AlGaN among the leading lateral power transistor platforms in the multi-kV regime, exceeding the state-of-the-art GaN and $Ga_2O_3$ devices in the same voltage range (Fig 6) [2-24].

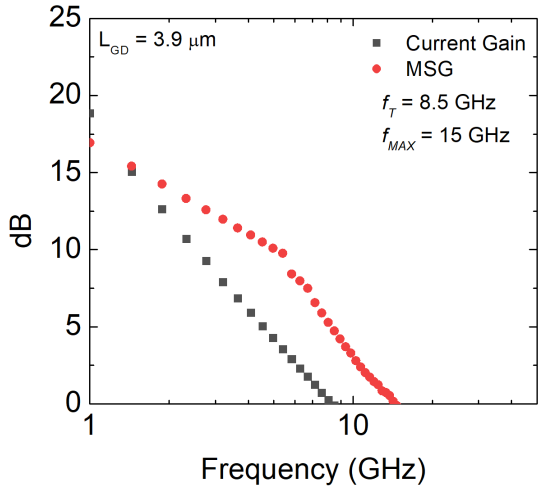
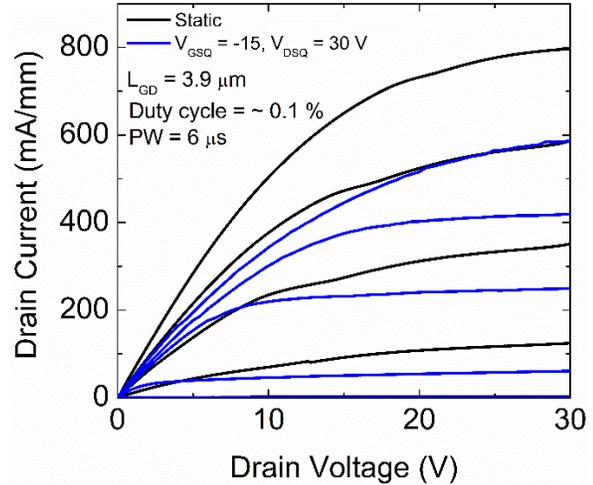

Figure 7. Small-signal measurements for $L_{GD}$ = 3.9 μm devices

Figure 8. Pulsed-IV measurement for $L_{GD}$ = 3.9 μm device, $\Delta V_G$ = -3 V, $V_G$ = 3 ~ -15 V

To evaluate the RF performance, on-wafer small-signal measurements were done using an Agilent 8510C vector network analyzer. Small-signal measurements were carried out at the DC bias point corresponding to the maximum transconductance. The extracted short-circuit current gain and maximum stable gain (MSG) for $L_{GD}$ = 3.9 μm are shown in Fig. 7. The cutoff frequency ($f_T$) and maximum oscillation frequency ($f_{MAX}$) were obtained to be 8.5 GHz and 15 GHz, respectively. These results indicate one of the highest reported combinations of $f_T$ (8.5 GHz) and $V_{BR}$ (1.28 kV) in UWBG AlGaN transistors, although the RF performance is primarily limited by relatively high contact resistance and the parasitic capacitance associated with the multiple field-plated structures. Excellent RF performance ($f_T$ = 85 GHz) in scaled device structure with similar epitaxial layers was reported in our previous work [1]. The trap-related effects were analyzed with pulsed I−V measurements using Keithley 4200 analyzer. The measurements were conducted on $L_{GD}$ = 3.9 µm devices at $V_{GSQ}$ = $V_p$ -2.5 V, and $V_{DSQ}$ = 30 V with 6 µs pulse width and 0.1 % duty cycle (Fig. 8). Current collapse suggests that $I_{MAX}$ under pulsed conditions was ~75% of $I_{MAX,DC}$.

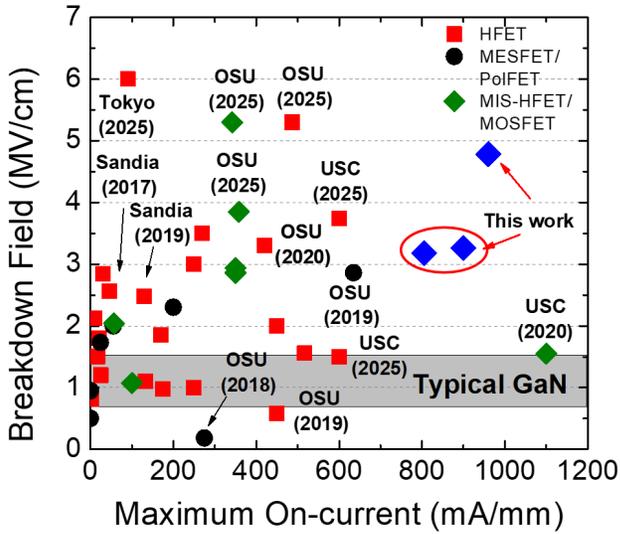

Figure 9. $F_{BR}$ versus $I_{MAX}$ benchmark plot in UWBG AlGaN transistors (channel Al % > 40 %) [19-48]

Fig. 9 shows a comparison of device performance and comparison among UWBG AlGaN transistors (channel Al composition > 40%) in terms of average breakdown field and maximum on-state current [19-48]. This benchmark indicates that PolFET structures in this work achieved a state-of-the-art combination of maximum on-state current and breakdown field in UWBG AlGaN, greatly exceeding previous work.

In conclusion, a UWBG AlGaN PolFET highlighting a high breakdown field (> 4.8 MV/cm) and excellent maximum on-state current (~ 960 mA/mm) was demonstrated. By utilizing gate-connected field-plated structures, a multi-kV breakdown performance was also achieved in longer $L_{GD}$ devices (~ 6.8 µm) with a low specific on-resistance, exhibiting 1.65 GW/cm² BFOM. Furthermore, the RF characteristics were investigated, exhibiting 8.5 GHz and 15 GHz for $f_T$ and $f_{MAX}$, respectively, in $L_{GD}$ = 3.9 µm devices. Pulsed I-V measurements indicated acceptable trap-related effects and current collapse, considering the use of PECVD $SiN_x$ passivation. The reported results indicate a state-of-the-art combination of $F_{BR}$ and $I_{MAX}$ in UWBG AlGaN transistors, highlighting the potential of UWBG AlGaN for both high-voltage switching and high-power RF applications

This work was funded by ARO DEVCOM under Grant No. W911NF2220163 (UWBG RF Center, program manager Dr. Tom Oder). This article has been co-authored by employees of National Technology & Engineering Solutions of Sandia, LLC under Contract No. DE-NA0003525 with the U.S. Department of Energy (DOE). The employee owns all right, title and interest in and to the article and is solely responsible for its contents. The United States Government retains and the publisher, by accepting the article for publication, acknowledges that the United States Government retains a non-exclusive, paid-up, irrevocable, world-wide license to publish or reproduce the published form of this article or allow others to do so, for United States Government purposes. The DOE will provide public access to these results of federally sponsored research in accordance with the DOE Public Access Plan https://www.energy.gov/downloads/doe-public-access-plan. SNL is managed and operated by NTESS under DOE NNSA contract DE-NA0003525

**Author Declarations**

**Conflict of Interest**

The authors have no conflicts to disclose.

**Data Availability**

The data that support the findings of this study are available within the article

**Supplementary materials**

The supplementary material includes buffer leakage comparisons between Si δ-doped and 2 nm uniformly doped n++ AlGaN back-barrier structures, as well as breakdown improvements enabled by gate-connected field-plate structures and the LGD-dependent breakdown field trends.

**References**

[1] Y. Zhu, A. A. Allerman, A. Wissel-Garcia, S. Shin, J. Pratt, C. Cao, K. J. Liddy, J. S. Speck, B. A. Klein, A. Armstrong, and S. Rajan, "Ultra-Wide Bandgap AlGaN Heterostructure Field Effect Transistors with Current Gain Cutoff Frequency Above 85 GHz (arXiv:2512.18103)," arXiv (2025).
[2] H. Jiang, Q. Lyu, R. Zhu, P. Xiang, K. Cheng, and K. M. Lau, "1300 V Normally-OFF p-GaN Gate HEMTs on Si With High ON-State Drain Current," IEEE Trans. Electron Devices 68(2), 653–657 (2021).
[3] R. Hao, W. Li, K. Fu, G. Yu, L. Song, J. Yuan, J. Li, X. Deng, X. Zhang, Q. Zhou, Y. Fan, W. Shi, Y. Cai, X. Zhang, and B. Zhang, "Breakdown Enhancement and Current Collapse Suppression by High-Resistivity GaN Cap Layer in Normally-Off AlGaN/GaN HEMTs," IEEE Electron Device Lett. 38(11), 1567–1570 (2017).
[4] X. Li, J. Zhang, J. Ji, Z. Cheng, J. Wang, L. Chen, L. Wang, S. You, L. Zhai, Q. Li, Y. Zhang, T. Liu, Z. Li, Y. Hao, and J. Zhang, "Demonstration of >8-kV GaN HEMTs with CMOS-compatible manufacturing on 6-in sapphire substrates for medium-voltage applications," IEEE Trans. Electron Devices 71(6), 3989–3993 (2024).
[5] J. T. Kemmerling, R. Guan, M. Sadek, Y. Xiong, J. Song, S.-W. Han, S. Isukapati, W. Sung, and R. Chu, "GaN super-heterojunction FETs with 10-kV blocking and 3-kV dynamic switching," IEEE Trans. Electron Devices 71(2), 1153–1159 (2024).
[6] M. Yanagihara, Y. Uemoto, T. Ueda, T. Tanaka, and D. Ueda, "Recent advances in GaN transistors for future emerging applications," Phys. Status Solidi A 206(6), 1221–1227 (2009).
[7] Y. Guo, Y. Qin, M. Porter, Z. Yang, M. Xiao, Y. Wang, D. Popa, L. Efthymiou, C. Cheng, K. Cheng, I. Kravchenko, L. Shao, F. Udrea, and Y. Zhang, "10 kV E-mode GaN HEMT: Physics for breakdown voltage upscaling," Appl. Phys. Lett. 127(4), 042102 (2025).
[8] J. Cui, J. Wei, M. Wang, Y. Wu, J. Yang, T. Li, J. Yu, H. Yang, X. Yang, J. Wang, X. Liu, D. Ueda, and B. Shen, "6500-V E-mode Active-Passivation p-GaN Gate HEMT with Ultralow Dynamic RON," in Proc. IEEE Int. Electron Devices Meeting (IEDM), pp. 1–4 (2023).
[9] H.-S. Lee et al., IEEE Electron Device Lett. 33(7), 982 (2012).
[10] M. Xiao et al., in Proc. IEEE Int. Electron Devices Meeting (IEDM), p. 114 (2021).
[11] J. Wei et al., in Proc. IEEE Int. Electron Devices Meeting (IEDM), p. 225 (2015).
[12] Y. Lv, H. Liu, X. Zhou, Y. Wang, X. Song, Y. Cai, Q. Yan, C. Wang, S. Liang, J. Zhang, Z. Feng, H. Zhou, S.

Cai, and Y. Hao, "Lateral b-Ga2O3 MOSFETs with high power figure of merit of 277 MW/cm2," IEEE Electron Device Lett. 41(4), 537–540 (2020).

[13] A. Bhattacharyya, S. Sharma, F. Alema, P. Ranga, S. Roy, C. Peterson, G. Seryogin, A. Osinsky, U. Singisetti, and S. Krishnamoorthy, "4.4 kV b-Ga2O3 MESFETs with power figure of merit exceeding 100 MW cm2," Appl. Phys. Express 15(6), 061001 (2022).

[14] A. Bhattacharyya, S. Roy, P. Ranga, C. Peterson, and S. Krishnamoorthy, "High-mobility tri-gate b-Ga2O3 MESFETs with a power figure of merit over 0.9 GW/cm2," IEEE Electron Device Lett. 43(10), 1637–1640 (2022).

[15] C. Wang, Q. Yan, C. Su, S. Alghamdi, E. Ghandourah, Z. Liu, X. Feng, W. Zhang, K. Dang, Y. Wang, J. Wang, J. Zhang, H. Zhou, and Y. Hao, "Demonstration of the b-Ga2O3 MOS-JFETs with suppressed gate leakage current and large gate swing," IEEE Electron Device Lett. 44(3), 380–383 (2023).

[16] C. Wang, H. Gong, W. Lei, Y. Cai, Z. Hu, S. Xu, Z. Liu, Q. Feng, H. Zhou, J. Ye, J. Zhang, R. Zhang, and Y. Hao, "Demonstration of the p-NiOx/n-Ga2O3 heterojunction Gate FETs and diodes with BV2/Ron,sp figures of merit of 0.39 GW/cm2 and 1.38 GW/cm2," IEEE Electron Device Lett. 42(4), 485–488 (2021).

[17] N. K. Kalarickal, Z. Xia, H.-L. Huang, W. Moore, Y. Liu, M. Brenner, J. Hwang, and S. Rajan, "b-(Al0.18Ga0.82)2O3/Ga2O3 double heterojunction transistor with average field of 5.5 MV/cm," IEEE Electron Device Lett. 42(6), 899–902 (2021).

[18] N. K. Kalarickal, Z. Feng, A. F. M. Anhar Uddin Bhuiyan, Z. Xia, W. Moore, J. F. McGlone, A. R. Arehart, S. A. Ringel, H. Zhao, and S. Rajan, "Electrostatic engineering using extreme permittivity materials for ultra-wide bandgap semiconductor transistors," IEEE Trans. Electron Devices 68(1), 29–35 (2021).

[19] K. Gohel, S. Mukhopadhyay, R. I. Roya, S. Sanyal, M. T. Alam, J. Chen, R. Bai, G. Wang, S. Pasayat, and C. Gupta, ">2.7 kV Al0.65Ga0.35N channel HEMT on bulk AlN substrate with >400 MW/cm2 Baliga figure of merit," IEEE Electron Device Lett. 46, 2102 (2025).

[20] R. Maeda, K. Ueno, A. Kobayashi, and H. Fujioka, "AlN/Al0.5Ga0.5N HEMTs with heavily Si-doped degenerate GaN contacts prepared via pulsed sputtering," Appl. Phys. Exp., vol. 15, no. 3, Mar. 2022, Art. no. 031002, doi: 10.35848/1882-0786/ac4fcf.

[21] M. T. Alam, J. Chen, K. Stephenson, M. A.-A. Mamun, A. A. M. Mazumder, S. S. Pasayat, A. Khan, and C. Gupta, "2 kV Al0.64Ga0.36N-channel high electron mobility transistors with passivation and field plates," Appl. Phys. Exp., vol. 18, no. 1, Jan. 2025, Art. no. 016504, doi: 10.35848/1882-0786/ad9db4.

[22] H. Tokuda, M. Hatano, N. Yafune, S. Hashimoto, K. Akita, Y. Yamamoto, and M. Kuzuhara, "High Al composition AlGaN-channel high-electron-mobility transistor on AlN substrate," Appl. Phys. Exp., vol. 3, no. 12, Dec. 2010, Art. no. 121003, doi: 10.1143/apex.3.121003.

[23] S. Shin, H. Pal, J. Pratt, J. Niroula, Y. Zhu, C. Joishi, B. A. Klein, A. Armstrong, A. A. Allerman, T. Palacios, and S. Rajan, "High breakdown electric field (>5 MV/cm) in UWBG AlGaN transistors," APL Electron. Devices 1(3), 036120 (2025).

[24] S. Shin, C. Cao, J. Pratt, Y. Zhu, B. A. Klein, A. Armstrong, A. A. Allerman, and S. Rajan, "Barrier electrostatics and contact engineering for ultra-wide bandgap AlGaN HFETs," APL Electron. Devices 1(4), 046131 (2025).

[25] J. Chen, P. Seshadri, K. Stephenson, M. A. Mamun, R. Bai, Z. Wang, A. Khan, and C. Gupta, "64% AlGaN channel HFET with high Johnson's figure of merit (>6THz V)," IEEE Electron Device Lett. 46, 545 (2025).

[26] R. Bai, S. Mukhopadhyay, K. Gohel, S. Sanyal, J. Chen, M. T. Alam, S. Xie, S. S. Pasayat, and C. Gupta, "Demonstration of high Johnson's figure of merit (ft × VBR > 20 THz V) and fmax × VBR (>42 THz V) for Al0.66Ga0.34N channel MISHEMT on bulk AlN substrates," Appl. Phys. Express 18(8), 086501 (2025).

[27] S. Mukhopadhyay, K. Gohel, S. Sanyal, M. Dangi, R. I. Roya, R. Bai, J. Chen, Q. Lin, G. Wang, C. Gupta, and S. S. Pasayat, "Characteristics of transport properties in ultra-wide bandgap Al0.65Ga0.35N channel HEMTs with low contact resistance and high breakdown voltage (>2.5 kV)," Appl. Phys. Lett. 126(15), 152103 (2025).

[28] S. Muhtadi, S. M. Hwang, A. Coleman, F. Asif, G. Simin, M. Chandrashekhar, and A. Khan, "High electron mobility transistors with Al0.65Ga0.35N channel layers on thick AlN/sapphire templates," IEEE Electron Device Lett. 38(7), 914–917 (2017).

[29] A. G. Baca, B. A. Klein, J. R. Wendt, S. M. Lepkowski, C. D. Nordquist, A. M. Armstrong, A. A. Allerman, E. A. Douglas, and R. J. Kaplar, "RF performance of Al0.85Ga0.15N/Al0.70Ga0.30N high electron mobility transistors with 80-nm gates," IEEE Electron Device Lett. 40(1), 17–20 (2019).

[30] A. G. Baca, B. A. Klein, A. A. Allerman, A. M. Armstrong, E. A. Douglas, C. A. Stephenson, T. R. Fortune,

and R. J. Kaplar, "Al0.85Ga0.15N/Al0.70Ga0.30N high electron mobility transistors with Schottky gates and large on/off current ratio over temperature," ECS J. Solid State Sci. Technol. 6(12), Q161 (2017).

[31] H. Okumura, S. Suihkonen, J. Lemettinen, A. Uedono, Y. Zhang, D. Piedra, and T. Palacios, "AlN metal–semiconductor field-effect transistors using Si-ion implantation," Jpn. J. Appl. Phys. 57(4S), 04FR11 (2018).

[32] H. Xue, C. H. Lee, K. Hussian, T. Razzak, M. Abdullah, Z. Xia, S. H. Sohel, A. Khan, S. Rajan, and W. Lu, "Al0.75Ga0.25N/Al0.6Ga0.4N heterojunction field effect transistor with fT of 40 GHz," Appl. Phys. Express 12(6), 066502 (2019).

[33] S. Bajaj, A. Allerman, A. Armstrong, T. Razzak, V. Talesara, W. Sun, S. H. Sohel, Y. Zhang, W. Lu, A. R. Arehart, F. Akyol, and S. Rajan, "High Al-content AlGaN transistor with 0.5 A/mm current density and lateral breakdown field exceeding 3.6 MV/cm," IEEE Electron Device Lett. 39(2), 256–259 (2018).

[34] A. M. Armstrong, B. A. Klein, A. G. Baca, A. A. Allerman, E. A. Douglas, A. Colon, V. M. Abate, and T. R. Fortune, "AlGaN polarization-doped field effect transistor with compositionally graded channel from Al0.6Ga0.4N to AlN," Appl. Phys. Lett. 114(5), 052103 (2019).

[35] S. Bajaj, F. Akyol, S. Krishnamoorthy, Y. Zhang, and S. Rajan, "AlGaN channel field effect transistors with graded heterostructure ohmic contacts," Appl. Phys. Lett. 109(13), 133508 (2016).

[36] J. Singhal, E. Kim, A. Hickman, R. Chaudhuri, Y. Cho, H. G. Xing, and D. Jena, "AlN/AlGaN/AlN quantum well channel HEMTs," Appl. Phys. Lett. 122(22), 222106 (2023).

[37] M. Gaevski, S. Mollah, K. Hussain, J. Letton, A. Mamun, M. U. Jewel, M. Chandrashekhar, G. Simin, and A. Khan, "Ultrawide bandgap AlxGa1−xN channel heterostructure field transistors with drain currents exceeding 1.3 A mm−1," Appl. Phys. Express 13(9), 094002 (2020).

[38] T. Razzak, S. Hwang, A. Coleman, S. Bajaj, H. Xue, Y. Zhang, Z. Jamal-Eddine, S. h. Sohel, W. Lu, A. Khan, and S. Rajan, "RF operation in graded AlxGa1−xN (x = 0.65 to 0.82) channel transistors," Electron. Lett. 54(23), 1351–1353 (2018).

[39] A. M. Armstrong, B. A. Klein, A. Colon, A. A. Allerman, E. A. Douglas, A. G. Baca, T. R. Fortune, V. M. Abate, S. Bajaj, and S. Rajan, "Ultra-wide band gap AlGaN polarization-doped field effect transistor," Jpn. J. Appl. Phys. 57(7), 074103 (2018).

[40] D. Khachariya, S. Mita, P. Reddy, S. Dangi, J. H. Dycus, P. Bagheri, M. H. Breckenridge, R. Sengupta, S. Rathkanthiwar, R. Kirste, E. Kohn, Z. Sitar, R. Collazo, and S. Pavlidis, "Record >10 MV/cm mesa breakdown fields in Al0.85Ga0.15N/Al0.6Ga0.4N high electron mobility transistors on native AlN substrates," Appl. Phys. Lett. 120(17), 172106 (2022).

[41] B. Da, D. Herath Mudiyanselage, D. Wang, Z. He, and H. Fu, "High-voltage kV-class AlN metal-semiconductor field-effect transistors on single-crystal AlN substrates," Appl. Phys. Express 17(10), 104002 (2024).

[42] Y. Kometani, T. Kawaide, S. Tanaka, T. Egawa, and M. Miyoshi, "AlN/AlGaN heterojunction field-effect transistors with a high-AlN-mole-fraction Al0.72Ga0.28N channel grown on single-crystal AlN substrate by metalorganic chemical vapor deposition," Jpn. J. Appl. Phys. 63(11), 111003 (2024).

[43] I. Abid, J. Mehta, Y. Cordier, J. Derluyn, S. Degroote, H. Miyake, and F. Medjdoub, "AlGaN channel high electron mobility transistors with regrown ohmic contacts," Electronics 10(6), 635 (2021).

[44] A. G. Baca, A. M. Armstrong, A. A. Allerman, E. A. Douglas, C. A. Sanchez, M. P. King, M. E. Coltrin, T. R. Fortune, and R. J. Kaplar, "An AlN/Al0.85Ga0.15N high electron mobility transistor," Appl. Phys. Lett. 109(3), 033509 (2016).

[45] J. Chen, K. Stephenson, M. A. Mamun, Z. Wang, P. Seshadri, A. Khan, and C. Gupta, "Al0.87Ga0.13N/Al0.64Ga0.36N HFET with fT > 17 GHz and Vbr > 360 V," in 2024 Device Research Conference (DRC) (IEEE, 2024), pp. 1–2.

[46] A. Mamun, K. Hussain, R. Floyd, M. D. Alam, M. Chandrashekhar, G. Simin, and A. Khan, "Al0.64Ga0.36N channel MOSHFET on single crystal bulk AlN substrate," Appl. Phys. Express 16(6), 061001 (2023).

[47] K. Ueno, R. Maeda, T. Kozaka, and H. Fujioka, "Degenerate GaN source–drain AlN/AlxGa1−xN/AlN high electron mobility transistors with a high breakdown electric field reaching 6.0 MV/cm," APL Mater. 13(4), 041129 (2025).

[48] J. Chen, A. Al Mamun Mazumder, P. Seshadri, D. Nandakumar, R. Bai, R. A. Choudhury, A. Khan, and C. Gupta, "Temperature dependent high frequency performance of a 62% AlGaN channel HEMT," APL Electron. Devices 2(1), 016102 (2026).